# PRODUCT ECODESIGN AND MATERIALS: CURRENT STATUS AND FUTURE PROSPECTS

F. Mathieux, D. Brissaud, P. Zwolinski
University of Grenoble
Grenoble, France

**ABSTRACT**

The aim of this paper is to discuss the current status of ecodesign in the industry and its future implications for materials.

There is today more and more focus on the environmental impacts of products during their whole life cycle. In particular, ecodesign aims at integrating environmental aspects during the product's design process as any other criterion, in order to reduce the life cycle impacts. Due to legislative pressure, customer requirements or even manufacturer's environmental policy, ecodesign is currently gaining in popularity in all industrial sectors. Although a lot of product environmental impact assessment and Design for Environment tools already exist, environmental aspects are unfortunately rarely routinely integrated into product development process in the industry. This is mainly due to the fact that current ecodesign tools are little adapted to designers' practices, requirements and competencies. After the sequential and DfX paradigms, design of products is today maturing into Integrated Design, where multiple points of views and expertise have to be considered at the same time to progressively define the product. For example, when choosing an acceptable material for a part, not only functional (e.g. aesthetic) and mechanical (e.g. strength) characteristics should be complied by this material, but its machinability, its recyclability and other factors will be important. In the near future, ecodesign of products will also have to be adapted to new productive paradigms such as Product / Service / System and Remanufacturing. These paradigms are indeed developing, obviously not only for environmental reasons, and ecodesign should be robust to these developments. At last, but not least, there is a real need for developing innovative tools and methods so that they can be used in the earlier phases of the product's design, so that real innovation and environmental benefits can be achieved.

## INTRODUCTION

The aim of this paper is to present our vision of what is the current status of products ecodesign in the industry and what might be its future implications for materials in the future.

There is today more and more focus on the environmental impacts of products and materials during their whole life cycle, i.e. not only during the production stage, but also during extraction of raw materials, distribution, use and end-of-life. In particular, ecodesign, that aims at integrating environmental aspects during the product's design process as any other criterion has been strongly developing and maturing. Today, although it is quite a well-known concept in the manufacturing industry, it is not widely and routinely implemented and it should be further elaborated to reach its full potential.

After a short history of product ecodesign and of its drivers, a brief analysis of its current implementation in the industry is presented (Section 1). Through the example of Design for Recycling, Section 2 discusses the necessary consolidation of ecodesign practices, when defining recognized practices, considering it within the larger context of integrated design, and mainstreaming into classical design tools. Section 3 focuses on new and promising manufacturing paradigms such as remanufacturing and Product/Service/Systems and on their implication on product ecodesign and material choices. Section 4 underlines the necessity to better integrate environmental aspects into early phases of the product development process.

## A SHORT HISTORY OF PRODUCT ECODESIGN AND CURRENT STATUS

### From end-of-pipe approach to product-centered prevention

The protection of the environment has been an increasing preoccupation of manufacturing industries since the seventies, due in particular to resources crisis, acute pollution events, or wider political strategies (e.g. the sustainable development principles enounced by the Brundtland Commission). From an industrial perspective, the traditional "end-of-pipe approach" of the seventies aiming at treating liquid, solid and gaseous effluents, has expended during the eighties and nineties into a more preventative approach, called "middle-of-pipe": it consists for example of minimizing waste and energy consumption. During the nineties, this manufacturing-centered pollution prevention approach has enlarged to the whole life cycle of products and materials through the development of products ecodesign. Ecodesign, as defined in ISO 14062, aims at integrating environmental aspects into product design and development, as any other criterion (quality, cost, security, time to market, etc.) and as soon as possible during the design process [1]. Although it is not the only parameter to be considered in ecodesign, material selection obviously plays an important role in the development of environmentally-conscious products. We describe below two major drivers of the development of ecodesign in the industry: legislation and customers expectations.

### Legislation

Legislation is an important driver for ecodesign development as many product categories are today targeted by EU regulations. For example, packaging products, motor vehicles and electric and electronic equipment are today covered by EU directives.

In general, EU directives define common rules concerning technical issues (e.g. end-of-life treatment of equipment; recycling rate to be achieved), financial issues (e.g. who pays for end-of-life treatment) as well as responsibility issues. EU directives also include essential prevention dispositions that encourage ecodesign: for example EU directives include hazardous substances restrictions and promote "the design and production of equipment […] which take into account and facilitate dismantling and recovery, in particular the reuse and recycling" (article 4 of [2]).

### Customer expectations

Fulfillment of customer requirements is also often seen as a driver for the development of ecodesign: this is in particular true for business to business products. It is also more and more true for consumer products as ecodesign is a way to highlight the company environmental policy [3]. Several strategies can be adopted for companies to demonstrate environmental performances of its products. It can either apply to obtain a recognized ecolabel or develop and diffuse to its customers products environmental profiles following recognized processes like the Environmental Product Declaration.

### Current development of ecodesign in industry

Ecodesign was traditionally mainly developed in sectors targeted by stricter environmental regulations, i.e. packaging, automotive and electr(on)ic industry. However, ecodesign is being generalized in many -if not all- other industrial sectors such as building, aeronautic, railways, food and drink or textile. Also, while ecodesign was in the past mainly driven by large groups and multinationals, it can be noticed today that small and medium-size enterprises are defining their own ecodesign strategies.

Until recently, ecodesign was mainly conducted through pilot projects where the leaders were ecodesign experts [3]. Therefore, such projects have been led with limited interaction with the rest of the design team. In general, ecodesign is indeed little routinely integrated into the product development [4]. This is however slowly changing and environmental experts are today more and more routinely involved in the design team.

## TOWARDS CONSOLIDATED PRACTICES IN DFE: THE EXAMPLE OF DESIGN FOR RECYCLING

### Consolidated practices in DfR

Since the start of ecodesign initiatives, Design for Recycling (DfR), i.e. integrating end-of-life aspects into the design of products, has been a very popular strategy. This can be explained by the fact that the impact of products at their end-of-life can easily be understood by designers, who are not environmental experts. More importantly, this is due to the fact that some legislations (e.g. packaging waste, ELV, WEEE directives) focus on products' end-of-life and designers are forced to consider it. Also, designers feel concerned by the end-of-life of products probably because implications of DfR on the design do concern products attributes that they routinely manipulate, namely [5]:

- Attributes of materials, e.g. material type, density, price, etc.
- Attributes of fastening, e.g. number and types of fastening,
- Attributes of architecture, e.g. modularity, accessibility, etc..

Therefore, in the early nineties, many DfR methods and tools have been developed in research labs and industries. A thorough inventory and analysis of DfR methods and tools is presented in [5]. Initially, most DfR methods were Dismantling-Conscious Design (DCD) approaches, where the dismantlability of the product should be optimized. A tool such as Restar [6] or guidelines such as the ones presented in [7] focus on DCD. Later on, considering that the economic viability of dismantling operations is not always guaranteed, some argued that products should instead be orientated to shredding processes coupled with sorting processes. Therefore, some Shredding-Conscious Design (SCD) approaches such as separability tables [8] that help designers to choose appropriate combination of materials have been developed. However, such approach was built on initial and partial model of product shredding. Fortunately, in recent years, valuable researches such as [9] and [10] contributed to a much better understanding of the shredding and sorting processes. Therefore, recent trends in research have been aiming at actually mixing DCD and SCD approaches to allow both dismantling, for depollution and re-use of some parts, and shredding/separation for the remaining fractions. Researches such as [11] and [5] are in line with such an approach.

However, these researches are still too recent to have defined new consolidated DfR practices, which could include, for example, typical acceptable material associations. To our perspectives, consolidation, test and diffusion of robust DfR practices is one of the challenges of the next years in the ecodesign field.

### DfR within the Integrated Design paradigm

After the sequential and DfX paradigms, design of products is today maturing into Integrated Design, where multiple points of views and expertises have to be considered at the same time to progressively define the product [12]. For example, when choosing an acceptable material for a part, not only functional (e.g. aesthetic), economical and mechanical (e.g. strength) characteristics should be complied by this material, but its machinability, its recyclability and other factors will be important. In such a context, DfR practices will have in the future to be discussed in a larger context, and recycling expert will have to interfere and find synergies with other experts within the design team. Within the same context, there are still some questions to be answers on whether environmental experts should be systematically integrated into the design team, or designers should expand their competencies.

### The need to adapt ecodesign tools to designer needs and practices

As argued earlier, ecodesign has not been routinely practiced in design teams: this is probably due to the facts that current ecodesign tools are too much expert tools that are little adapted to designers' current needs, tools and practices. If environment and more particularly recycling is to be more considered as any other parameter in the design process, tools will have to be adapted. Therefore, it is necessary that existing environmental impact assessment and ecodesign tools are revisited, and if possible connected to engineers' design tools currently used in companies. Example of design tools such as Product Life cycle Management systems or Material Selection Tools should in particular be explored.

## A NECESSARY ADAPATION OF ECODESIGN TO NEW PRODUCTIVE PARADIGMS

The increasing mass of products coming to end of life becomes today a real problem for the environment in parallel with the increase of the energy and material costs. Moreover, with new environmental directives, manufacturers become responsible for the treatment of their products at the end of life. This leads manufacturers to reconsider the concept of property of the product and to develop new strategies to limit end of life treatment costs and material costs [13]. They can adopt two main approaches:

- The manufacturer pays an external firm to recover and recycle its products. In this case, the recycling will certainly be limited to the material recovery, because of the diversity of the products treated by the external recycling firms;
- The manufacturer recover the products himself to try to make a benefit while keeping more added value on his product than those obtained by the simple price of the recycled materials.

As seen earlier in this paper, the now classical scenarios of waste recovery as secondary materials must tackle technological and economic obstacles. From a technical point of view, the issue of the substitutability of secondary materials is crucial: introducing secondary materials into the production cycle depends on the physical and chemical properties of the materials under consideration. From an economic point of view, it is now accepted that the profitability of material recycling is relatively poor. Moreover, the high level of uncertainty on recycled material prices and the competition of original raw materials make investment in recycling systems very risky.

So, designing products, which are recovered by the manufacturer and "never" discarded, could be a real and beneficial future objective for manufacturers. Indeed, this will be a mean to keep more added value than if there is only a material recycling process. Moreover, this approach expands the product life cycle because (at least) a second life cycle will be defined for it (or for a part of it). In the following sections, we will present two new productive paradigms that are developed not only due to environmental constraints, but also for economical benefits for firms: remanufacturing and product service systems.

**The remanufacturing**

The remanufacturing is an end of life strategy that reduces the use of raw materials and saves energy while preserving the added-value during the design and manufacturing processes. The remanufacturing process aims at extending the life of products by diverting them to a second life instead of disposing them of. The remanufacturing process is a process in which reasonably high volumes of similar products are collected to a central service place, disassembled, and then treated to be reused [14]. Therefore by keeping the part shape, material extraction and energy consumption can be reduced. The remanufacturing process is generally composed of several stages: disassembly, testing, repair, cleaning, inspection, updating, component replacement and assembly [15] [16]. At each stage, specific measures guarantee quality control (Figure 1).

The economic interest comes from the fact that the added-value due to the initial production of the product is fully or partly preserved. The environmental interest comes from the lower resources consumption compared to manufacturing a second new product, and the extended product life. However, life cycle analyses have still to be realized to verify that environmental benefits during the production phase are not overcome by environmental losses during other phases (e.g. energy saving during the use phase brought by new technologies).

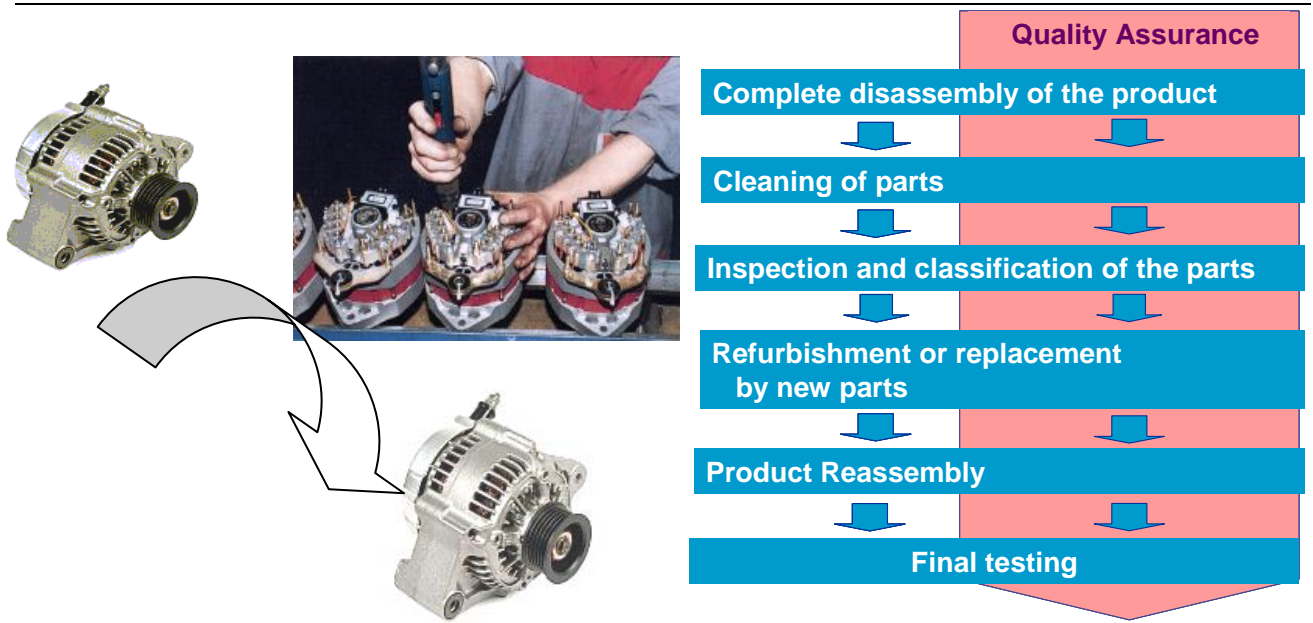

*Figure 1 The remanufacturing process*

Currently, in most of the cases, remanufacturing processes must be adapted to existing products because products have not been initially designed to be remanufacturable [17]. However, the process adaptations increase costs and this can lead the overall benefits of the remanufacturing process to be reconsidered. Therefore, new technical criteria have to be defined to design true remanufacturable products, to increase their performances and their life time. This also means that new specific materials should be developed with new performances (e.g. ability to be cleaned, life time increase).

**The Product Service System approach**

In developed countries, more and more enterprises are currently switching from selling physical products to providing service solutions. It enables companies to customize solutions and moreover to stand out from competitors. In addition, by providing service and not anymore selling physical products, it could be possible to decrease waste [18]. This awareness of innovation and sustainability leads to the concept of product-service system (PSS) [19]. PSS are seen as a new selling approach in which both physical products and immaterial services are gathered to fulfill customer's needs [20]. Consequently, the development of PSS can be achieved through service design or product design. But, products and services are involved in a global set, thus it is necessary to develop product and services within an integrated design process in order to optimize cost, quality and time-to-market [21]. Since products are not stand alone but included in a global system, some modifications must be done in order to make the system working efficiently and to be economically attractive [22].

For many companies, financial savings and revenues generated from shifting to services oriented-solutions is the most important driver. But there are also some researchers who state that environmental improvement is the main driver for the shift towards service oriented solutions [23], while increasing the potential for reuse, refurbishment, upgrading of products and recycling of materials. Indeed, if the example of a producer that owns the product all along its lifecycle is considered, it can be easily imagined that product updating, take back, remanufacturing and recycling processes will be more economic and easier to manage. In that case, life cycle strategies can ensure the reduction of material use in the production of new products and ensure a secondary source of raw materials directly from the market.

## ECODESIGN AND EARLY PHASES OF THE DESIGN

During the design of a product, this means during the complete technical development of a new product, different levels exists for the application of the eco-design: eco-optimization, eco-structure and eco-function:

- The eco-optimization: in this phase of the product development, the designer optimizes each subset of the product in order to minimize the potential environmental impacts throughout its lifecycle. For this purpose, he applies rules on well defined technical systems or on components of the product, in order to develop a characterization of the product related to the environment (e.g. choosing among materials more adapted to the life of the product, or to its end-of-life strategy). It is also at this level that the Life Cycle Analysis tools can be used. This phase addresses the detailed design of the product.

- The eco-structure definition: in this phase, the designer defines the overall structure and the functioning principles for a product allowing the reduction of the environmental impacts. Innovation comes while working on the functioning principles of the product (technical functions -> choice of solutions) and either on the architecture of the product. This level is related to the "conceptual design" and "embodiment design" phases during the design process. Very few tools exist for these phases: for example, materials are currently chosen only depending on technological criteria linked to the usage phase.

- The eco-function definition: In this phase, designers detect the appropriate use allowing the reduction of the environmental impacts only by the way in which the product is used. The most appropriate scenarios of use of the product are determined even if their intrinsic design features are not yet defined. The system of sale and the company's strategy are considered and designers reason on scenarios of use in order to define the main functions to be ensured by the product proposed (example: the product service system (PSS) for a Xerox photocopiers).

Some researches [24, 25] underlined that, the primary task of the conceptual design phase is to satisfy the functional requirements. However, it is important to consider the fact that, when aiming at satisfying the functional requirements, the designer must take decisions which significantly condition all aspects of a product. In particular, the costs are determined by the correctness of the operated choices. Nevertheless, this phase, unlike all other phases that compose the design process, is also the one which is less supported by dedicated tools.

For example, if we consider Dismantling-Conscious Design, it appears that no tools and no implemented methods exist to help the designer between the conceptual design phase and the detailed design phase (cf. Figure 2).

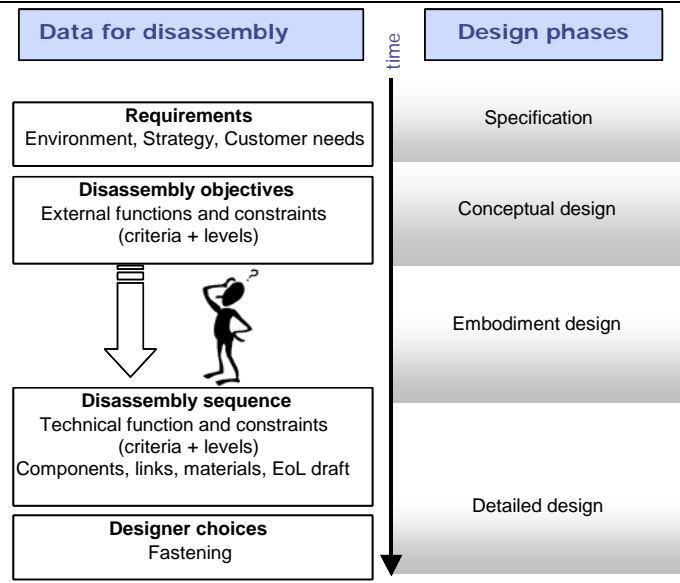

*Figure 2. Needs for Dismantling-Conscious Design tools between the conceptual design phase and the detailed phase.*

A similar analysis has been realized by Deng and al. on materials [26] that concludes that nothing exists to take into account environmental aspects of materials early during the design process. It is the reason why they have proposed to develop a model to bring back material consideration from the detailed design to the conceptual design stage.

## CONCLUSION

This paper presented our vision of the current status of product ecodesign in the industry, of its future trends and of its implications for materials. After defining product ecodesign and situating its historical context, it was shown that, although ecodesign is becoming more popular, it is still not routinely implemented in the industry. This is due to several facts. First, even within the ecodesign community, not all practices are fully consensual and some of them have still to be consolidated: this in particular true for material choices in Design for Recycling practices. Secondly, very few attempts focused on situating the environmental expertise within the larger context of product integrated design: material choices indeed impact on costs, manufacturing, recyclability, etc. Finally, ecodesign tools are still little adapted to designers needs and tools. Research in ecodesign should now tackle these problems. Moreover, as new productive paradigms such as product remanufacturing or Product/Service/System are developing, ecodesign and material selection should be adapted so that it fully integrates these new constraints. At last but not least, new ecodesign tools and methods should be developed to be usable in early design stages, so that real innovation and environmental benefits can be achieved. Ecodesign will only be robust and fully efficient when these challenges will be taken up.

## REFERENCES

1.  ISO, *Environmental management - Integrating environmental aspects into product design and development.* 2002, International Standard Organisation.

2. EU, *Directive of the European Parliament and of the Council on the Waste Electric and Electronic Equipment*. 2002, European Union: Brussels (Belgium).
3. Mathieux, F., Rebitzer,G., Ferrendier,S., Simon,M., Froelich,D., *Ecodesign in the European Electr(on)ics Industry - An analysis of the current practices based on cases studies.* Journal of Sustainable Product Design, 2001. **1**(4): p. 233-245.
4. Lindahl, M., *Engineering designers' experience of design for environment methods and tools - Requirements definitions from an interview study.* Journal of Cleaner Production, 2006. **14**: p. 487-496.
5. Mathieux, F., Froelich,D., Moszkowicz,P., *ReSICLED: a new Recovery-Conscious Design method for complex products based on a multicriteria assessment of the recoverability.* Journal of Cleaner Production, 2006.
6. Navin-Chandra, D. *ReStar: a design tool for environmental recovery analysis*. in Proceedings of *International Conference on Engineering Design*. 1993. The Hague (The Netherlands).
7. Simon, M., Mc Laren, J., Dowie, T., *Feasibility of a common indexing system for recycling of electronic products*. 1996, Manchester Metropolitan University: Manchester (United Kingdom). p. 19p.
8. AFNOR, *Automotive vehicles - Design of the vehicles that aim at the optimisation of their end-of-life recovery*. 1996, Association Française de Normalisation: Paris (France).
9. Aboussouan, L., Russo,P., Pons,M., Thomas,D., Birat,J.P., Leclerc,D., *Steel scrap fragmentation by shredders.* Powder Technology, 1999(105): p. pp.288-294.
10. van Schaik, A., Reuter,M.A., Heiskanen,K., *The infuence of particle size reduction and liberation on the recycling rate of end-of-life vehicles.* Minerals Engineering, 2004. **17**(2): p. 331-347.
11. Ferrão, P., Amaral,J., *Design for recycling in the automotive industry: new approaches and new tools.* Jorunal of Engineering Design, 2006. **17**(5): p. 447-462.
12. Tichkiewitch, S., Brissaud,D., *Methods and Tools for Cooperative and Integrated Design*. 2004, Kluwer Academic Publisher.
13. Zwolinski, P., Prudhomme, G., Brissaud, D., *Environment and Design. Methods and Tools for Integration and Co-operation*, in *Methods and Tools for Cooperative and Integrated Design*. 2003, Kluwer Academic Publishers. p. 11p.
14. Sundin E., J.N.B.M. *Analysis of Service Selling and design for remanufacturing*. in Proceedings of *IEEE international Symposium on Electronics and the Environment*. 2000. San Francisco, CA, USA.
15. Kerr, W., *Remanufacturing and eco-efficiency*. 1999, International Institute for Industrial Environmental Economics (IIIEE). Lund University: Lund (Sweden).
16. Sherwood, M., Shu, L., *Supporting design for remanufacture through waste-stream analysis of automotive remanufacturers.* CIRP Annals, 2000. **49**(1): p. pp. 87-90.
17. Zwolinski P., L.-O.M., Brissaud D., *Integrated design of remanufacturable products based on product profiles.* Journal of Cleaner Production, 2006. **14**(15-16): p. pp 1333-1345.
18. Manzini E., V.C., Garette C., *Product-Service Systems: Using an existing concept as a new approach to sustainability.* Journal of Design Research, 2001. **1**(2).
19. Tukker A., T.U., 2006, *New business for old Europe: Product-Service development, competitiveness and sustainability*. 2006, Sheffield (UK): Greenleaf Publishing.
20. Goedkoop M., v.H.C., de Riele H., Rommens P., *Product-Service System: ecological and economical basics*, 1999, PRé Consultants: Place Published. http://www.pre.nl/pss/download_PSSreport.htm (accessed 12/01/2006).

21. Maussang N., B.D., Zwolinski P. *Which representation for sets of products and associated services during the design process?* in Proceedings of *International Design and Manufacturing in Mechanical Engineering*. 2006. Grenoble (France).
22. Mont O., D.C., Jacobsson N. Journal of Cleaner Production, *A new business model for baby prams based on leasing and product remanufacturing.* Journal of Cleaner Production, 2006. **14**(17): p. pp. 1509-1518.
23. Maxwell D., S.W., van der Vorst R., *Functional and systems aspects of the sustainable product and service development approach for industry.* Journal of Cleaner Production, 2006. **14**(17): p. pp. 1466-1479.
24. Giampà, F., Muzzupappa, M., Rizzuti, S. *Design by function: a methodology to support designer creativity*. in Proceedings of *Proceedings of Design 2004*. 2004. Dubrovnik (Croatia).
25. Bruno, F., Giampà, F., Muzzupappa, M., Rizzuti, S. *A methodology to support designer creativity during the conceptual design phase of industrial products*. in Proceedings of *ICED03*. 2003. Stockholm (Sweden).
26. Deng, Y., Lu, W.F. *From function to structure and material: a conceptual design framework*. in Proceedings of *TMCE 2004*. 2004. Lausanne (Switzerland).